\documentclass[numberedappendix,iop]{emulateapj}
\usepackage{rotating}
\usepackage{epsfig}
\usepackage{bm}

\newcommand{\be}{\begin{eqnarray}}
\newcommand{\ee}{\end{eqnarray}}

\singlespace
\newcommand{\captionfonts}{\fontsize{10}{10}\selectfont}

\makeatletter  
\long\def\@makecaption#1#2{%
  \vskip\abovecaptionskip
  \sbox\@tempboxa{{\captionfonts #1: #2}}%
  \ifdim \wd\@tempboxa >\hsize
    {\captionfonts #1: #2\par}
  \else
    \hbox to\hsize{\hfil\box\@tempboxa\hfil}%
  \fi
  \vskip\belowcaptionskip}
\makeatother

\begin{document}

\title{Pulse intensity modulation and the timing stability of millisecond pulsars:  A case study of PSR~J1713$+$0747}
\shorttitle{The timing stability of  PSR~J1713$+$0747}
\shortauthors{Shannon \& Cordes}
\author{
  Ryan~M.~Shannon\altaffilmark{1,2} \& James.~M.~Cordes\altaffilmark{2}}
\altaffiltext{1}{CSIRO Astronomy and Space Science, Box 76, Epping, NSW 1710, Australia}
\altaffiltext{2}{Astronomy Department, Cornell University, Ithaca, NY 14853, USA}
\email{ryan.shannon@csiro.au}
\email{cordes@astro.cornell.edu}

\begin{abstract}
Most millisecond pulsars, like essentially all other radio pulsars, show timing errors well in excess of what is expected from additive radiometer noise alone.  We show that changes in amplitude, shape and pulse phase for the millisecond pulsar J1713$+$0747 cause this excess error. These changes appear to be uncorrelated from one pulse period to the next.   The resulting time of arrival variations are correlated across a wide frequency range and are observed with different backend processors on different days, confirming that they are intrinsic in origin and not an instrumental effect or caused by strongly frequency dependent interstellar scattering.  Centroids of single pulses show
an rms phase variation $\approx 40~\mu$s, which dominates the timing error and is the same phase jitter phenomenon long known in slower   spinning, canonical pulsars.  We show that the amplitude modulations of single pulses are modestly correlated with their arrival time fluctuations. We also demonstrate that single-pulse variations are completely consistent with arrival time variations of pulse profiles  obtained by integrating $N$ pulses such that the arrival time error decreases proportional to $1/\sqrt{N}$.  We investigate methods for correcting times of arrival for these pulse shape changes, including multi-component TOA fitting and principal component analysis. These techniques are not found to improve the timing precision of the observations.
  We conclude that when pulse shape changes dominate timing errors, the timing precision of PSR~J1713$+$0747 can  be only improved by averaging over a larger number of pulses.
 \end{abstract}

\section{Introduction}

The timing  of pulsars has enabled many studies of fundamental importance to physics and astrophysics, including precision tests of general relativity \cite[][]{1982ApJ...253..908T,2006Sci...314...97K},  strong constraints on nuclear equations of state \cite[][]{2007PhR...442..109L,2010Natur.467.1081D}, and the discovery of the first Earth-mass planets outside of the solar system \cite[][]{1992Natur.355..145W}.
By studying the correlated variations in pulse times of arrival (TOAs) from a set of millisecond pulsars (MSPs) that comprise a pulsar timing array (PTA),  
it is possible to detect  nanohertz gravitational waves (GWs) \cite[][]{1983ApJ...265L..39H,1990ApJ...361..300F}.
Candidate sources of GWs in the nanohertz frequency band are binary supermassive black holes in the throes of merger, oscillating cosmic strings, and the inflation era Universe \cite[][]{2009arXiv0909.1058J}. 
The net perturbation due to these sources to TOAs is expected to be at most tens of nanoseconds over $5-10$ years.

The most plausible source of gravitational waves is a stochastic background associated with massive black hole binaries \cite[][]{2003ApJ...583..616J,2008MNRAS.390..192S}. 
 For this background, a modestly significant  ($\lesssim 4\sigma$) detection can be achieved by observing $20$~to~$40$~pulsars over $5$~to~$10$~years \cite[][]{2005ApJ...625L.123J} assuming monthly observations with $100$~ns timing precision.

The required timing stability of $\lesssim 100$~ns can only be achieved 
with MSPs, pulsars with spin periods between $\approx 1$ and 
$20$~ms, which are more spin stable than canonical pulsars (CPs, pulsars with spin periods between $\approx 30$~ms and $8$~s). 
Over periods of years, CPs  show large levels of correlated spin noise likely associated with instabilities in both the pulsar magnetosphere and the neutron star itself \cite[][]{1985ApJS...59..343C,1995MNRAS.277.1033D,2010Sci...329..408L,sc2010}.

The precision of pulsar timing is at minimum limited by radiometer noise associated with the telescope receiving system and sky background.
 If this is the only form of noise, the sensitivity of observations is improved by increasing observation bandwidth, increasing observing time, or by increasing telescope sensitivity through larger collecting area or lower
system temperature.   

 However, there are certainly other processes that affect the timing precision of pulsars and limit the sensitivity of PTAs to GWs.  The sources and importance of a wide range of effects 
are summarized in \cite{cs2010}.   
Stochastic timing variations can  be broadly  bifurcated into those with
stationary statistics, such as white noise, and others that have
or appear to have nonstationary statistics, such as those with very broad
fluctuation spectra, including steep power laws with most of the  power at
low frequencies, often described as ``red'' processes. 
Red-like timing noise appears to describe the residuals seen in many
canonical pulsars though it is still under debate whether on the longest times scale,
 the timing variations have non-stationary
statistics or show band-limited or quasi-periodic statistics   \cite[][]{2010Sci...329..408L,sc2010}. 
Propagation of radio emission through the interstellar medium \cite[][]{1984Natur.307..527A,2010ApJ...717.1206C,cs2010}  also produces timing variations that
have red power spectra.      

On shorter time scales ranging from one pulse period to integrations from
several minutes to an hour,
TOA variations
exceed what is expected from radiometer noise alone.
In a study of CPs, \citet[][]{1985ApJS...59..343C}  
demonstrated that timing errors were generally always larger than expected
from additive radiometer noise and interpreted the excess in terms of
changes in pulse shape and amplitude
on pulse-to-pulse time scales. Termed ``jitter,'' this phenomenon has been seen in virtually
all pulsars that have been closely investigated.
Integrated pulse
profiles include these variations and thus so do the TOAs derived from
them.   


While the link between single pulses and timing variations is well
established in CPs, studies of MSPs are fewer because MSPs typically have much lower flux densities.
However, a few cases indicate  that single-pulse variations 
in MSPs  are consistent with those seen in CPs.
\cite{1998ApJ...498..365J} studied the variability of single pulses of the bright millisecond pulsar J0437$-$4715. They found that  single pulses contained a wide variety of morphologies and concluded the emission  showed similar properties to that observed in canonical pulsars.   \cite{2011arXiv1108.0812O} followed up this analysis by identifying a correlation between arrival time and pulse shape changes, and used this correlation to improve timing precision by $20\%$.
In a more recent study of PSR~J0437$-$4715, \cite{2012MNRAS.420..361L}   found excess timing errors that they attributed to pulse shape variations. 
  \cite{2003A&A...407..273E} analyzed the intensity modulation of a set of northern MSPs and suggested that there are periodic intensity modulations in MSP arrival times comparable to the drifting sub-pulse phenomenon observed in many canonical pulsars.
\cite{2004ApJ...602L..89J} found remarkable intensity stability of the main pulse of PSR~B1937$+$21, but accompanied by 
large intensity modulation in giant pulse emission.
\cite{1999ApJ...520..324K} showed that the relative intensity of the components of the MSP J1022$+$1001  
change on longer time scales and that a better timing solution for this pulsar could be obtained when these changes were incorporated.

In this paper, we focus on high S/N observations of the millisecond pulsar J1713$+$0747 to analyze fast pulse shape variations through 
measurements of single pulses and average pulse profiles.
In Section \ref{sec:observations_analysis_1713}, we summarize the observations and the data reduction procedure.   
In Section \ref{sec:jitter_1713}, we infer both directly and statistically  that intrinsic pulse shape changes cause TOA variations in this pulsar.  
 In Section \ref{sec:confirm_jitter}, we rule out other processes that mimic pulse jitter in our observations. 
 In Section \ref{sec:correction}, we present two different correction procedures that can be used to correct for TOA variations and apply the techniques (unsuccessfully) to our observations.  
In Section \ref{sec:discuss}, we discuss observing strategies that mitigate the effects of pulse profile variations on timing precision.

\section{Observation and Analysis Procedure} \label{sec:observations_analysis_1713}



The relatively nearby MSP J1713+0747 has a period $P=4.57$~ms, dispersion
measure DM$ =16$~pc~cm$^{-3}$, and a pulse width of 110$~\mu s$ (FWHM) \cite[][]{2005AJ....129.1993M}.
Its large (period averaged) flux density of 7.4~mJy at 1.4~GHz  contributes
to its being one of the best MSPs in terms of timing precision.
We observed PSR~J1713$+$0747 at 1.5~GHz with the Arecibo 
$305$-m telescope\footnote{The Arecibo Observatory is part of the National Astronomy and Ionosphere Center, which was operated by Cornell University 
under a cooperative agreement with the National Science Foundation.} 
using the ``L-band Wide'' receiver, 
which provides two channels, one for each hand of circular polarization. 
The results presented here use data recorded with two 
backend signal processing instruments:  the Wide-band Arecibo Pulsar Processor  \cite[WAPPs, ][]{2000ASPC..202..275D}  and 
the Arecibo Signal Processor \cite[ASP, ][]{2007PhDT........14D} .   
The ASP backend is used for several long-term timing programs, including 
precision timing for GW detection.   We use the WAPP data to study 
single-pulses but the ASP data to analyze TOA and pulse-shape variations 
of average profiles. The combination of data then allows us
to  connect single-pulse
results to the long-term timing programs on PSR~J1713+0747  that make use
only of average profiles. 

The WAPP data provided $\sim 50k$ pulses over a 4-minute time span on 
MJD 54983 and were used to analyze single pulses and 
sub-integrations of up to 512 pulses.  
Specifically, the WAPP instrument outputs a time series of
 autocorrelation functions (ACFs) of both  polarizations 
with $3$-level sampling.    
Offline, the {\tt sigproc} package\footnote{\tt sigproc.sourceforge.net} 
was used to Fourier transform the ACFs into spectra
with  $192$~frequency channels across a  $100$~MHz total bandwidth 
spanning $1.57$~to~$1.67$~GHz 
with $32~\mu$s~resolution.  After excising RFI, the data were dedispersed 
to form a single time series in the sum of the two polarization channels, which we call the total intensity.  
We used the {\tt dspsr} software package \cite[][]{2011PASA...28....1V}  to fold the data into subintegrations containing $1$~to~$1000$ pulses, and the {\tt psrchive} software package \cite[][]{2004PASA...21..302H} to measure TOAs from the subintegrations. 

The ASP backend removes dispersion coherently 
to produce dedispersed time series in all four Stokes parameters which
were then synchronously averaged in real time using the pulsar ephemeris.
Averages were outputted every  
$10$~s for 16 sub-bands, each  with $4$~MHz bandwidth, for a total
of  $64$~MHz centered at $1410$~MHz.
The {\tt ASPFitsReader} pipeline was used to calibrate and generate TOAs \cite[][]{2008PhDTFermdan} from ASP measurements.  
Polarization calibration and absolute flux calibration were completed by comparing a pulsed signal generated by a noise diode while the telescope was pointed on and off a calibrating source to the pulsed signal while the telescope was pointed at the pulsar.     
The radio galaxy CTD~93 was used as the calibration source 
because it is known to be both unpolarized and to have no observed 
flux variability \cite[][]{1999ApJ...515..558S}.
For each channel, the receiver was calibrated for differential gain and  phase offset between the feeds.   

We calculated TOAs from both the  WAPP and ASP data
by using a matched-filter algorithm implemented in the Fourier domain
\citep{1992RSPTA.341..117T}.
In this approach, TOAs are estimated  by comparing the observed total intensity pulse profile $I(\phi)$ to a standard template $T(\phi)$.    
The matched-filtering approach is based on the assumption that the 
measured pulse profile is a scaled version of the template added to
a measurement error term,
\be
\label{eqn:TOA_eqn}
I(\phi) = a T(\phi-\phi_o) + b + n(\phi),
\ee
where $a$, $b$ and $\phi_o$ are unknown parameters, and $n(\phi)$ is noise.  
Under the assumptions underlying Equation (\ref{eqn:TOA_eqn}), the 
TOA uncertainty is $P \delta \phi_o$, where $\delta \phi_o$ is the standard
error of $\phi_o$.

For the ASP observations, templates for each observing band were generated from high-S/N observations of the pulsar co-added from the long-term Arecibo/NANOGrav\footnote{\tt www.nanograv.org}  timing program, which at present uses 
backend instrumentation and observing bands  identical to those used
in  the work presented here (M.~Gonzalez, private communication).  For 
the WAPP observations, we produced an analytic template containing 
five Gaussian components,  consistent with the average profiles 
in these observations, and comparable to a previous model of the 
profile described in \cite{1998ApJ...501..270K}.

Residual TOAs were then calculated by fitting a timing model ephemeris to the TOAs.  For the observations here, we used an initial timing ephemeris derived from long term monitoring of J1713+0747 conducted at the Arecibo observatory \cite[][]{2012arXiv1201.6641D}.



\section{Evidence for Pulse Phase Jitter}\label{sec:jitter_1713}

\subsection{Single Pulse Profile Variations}\label{sec:singlepulse}

\begin{figure}[!ht]
\begin{center}
\includegraphics[scale=0.5]{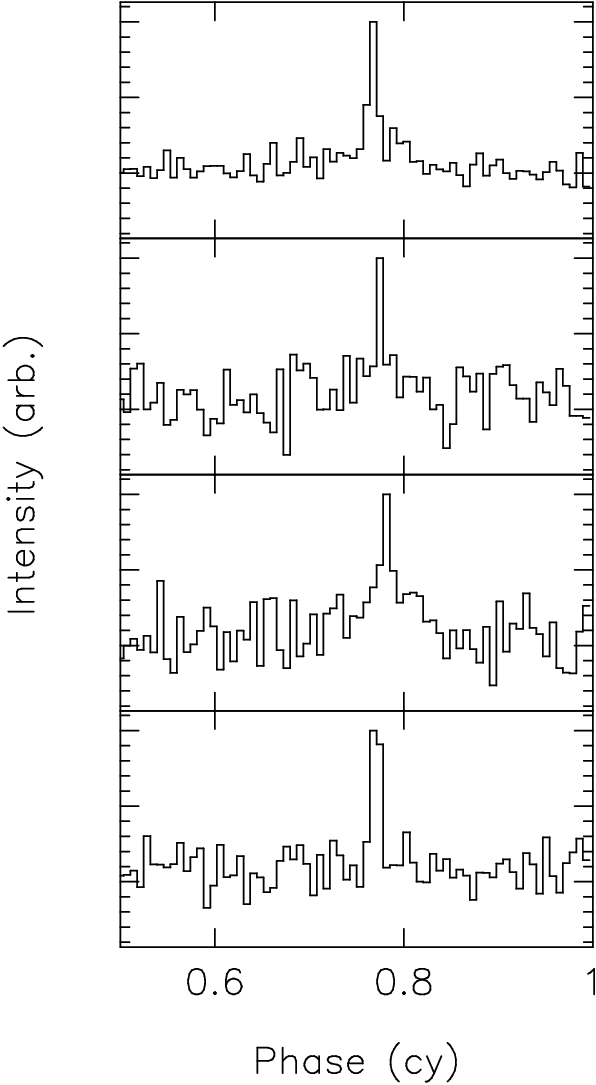}
\caption{ \label{fig:pulses} Pulse intensity versus phase (measured in cycles, cy) for four single pulses from PSR~J1713$+$0747.   The pulses have been scaled to the same peak amplitude to show differences in shape and centroid pulse phase.   
}
\end{center}
\end{figure}

In this section, we examine single pulses from  PSR~J1713$+$0747. 
The 50$k$ individual pulses obtained over four minutes allow
pulse shape variations to be characterized during an interval
when the modulation from diffractive interstellar scintillation
(DISS) is approximately constant, since the  characteristic DISS  time
scale is about one hour at 1.5~GHz.  

The single pulse signal to noise ratio ($S/N$, defined as the ratio of peak intensity to the rms off-pulse intensity) ranged from approximately $1$ to $30$.      
In Figure \ref{fig:pulses}, we show four relatively bright single pulses 
that have different morphologies with peak locations and 
pulse widths varying by amounts that cannot be attributed to radiometer noise. 

\begin{figure}[!ht]
\begin{center}
\includegraphics[scale=0.5]{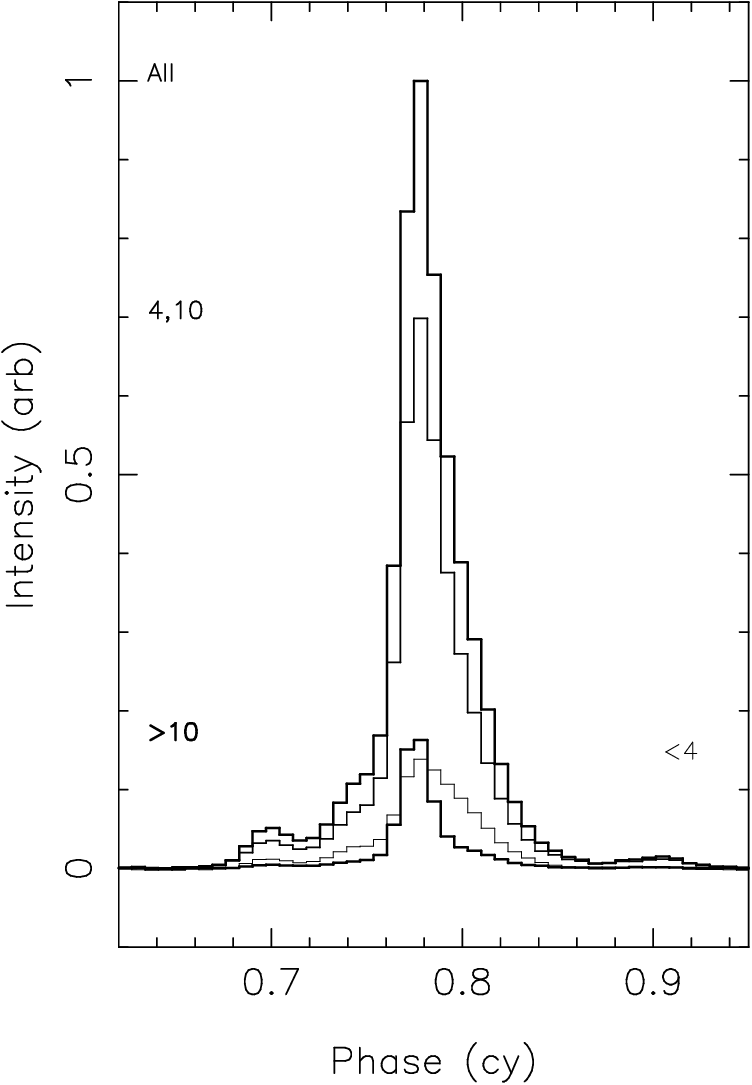}
\caption{ \label{fig:sn_budget} Average pulse profiles for different ranges of $S/N$.
The pulse profiles were formed from single pulses with $S/N$ in the different ranges.  
The ranges are $S/N<4$, $4<S/N<10$ (labeled on the plot as $4,10$), $S/N>10$, and  for all $S/N$ ranges, and are labeled on the plot at a height close to the peak flux for each  pulse profile, in a line thickness corresponding the thickness of the pulse profile.     The profiles are normalized to the total fluence so the profiles reflect the relative flux being emitted in these $S/N$ ranges.  
}

\end{center}
\end{figure}

Figure \ref{fig:sn_budget} shows average profiles 
that have been calculated using single pulses selected from different 
$S/N$ ranges; each profile has been normalized to reflect the 
total detected flux in the S/N range. 
Despite occurring three times less frequently than the faintest pulses, 
the brightest and narrowest pulses contribute a total flux that is comparable
to that of  the faintest pulses. 
We also find that brighter pulses are narrower with peaks
concentrated closer to the leading edge of the average pulse, 
suggesting that brighter pulses 
have earlier arrival times, which we will show below.

\begin{figure}[!ht]
\begin{center}
\includegraphics[scale=0.5]{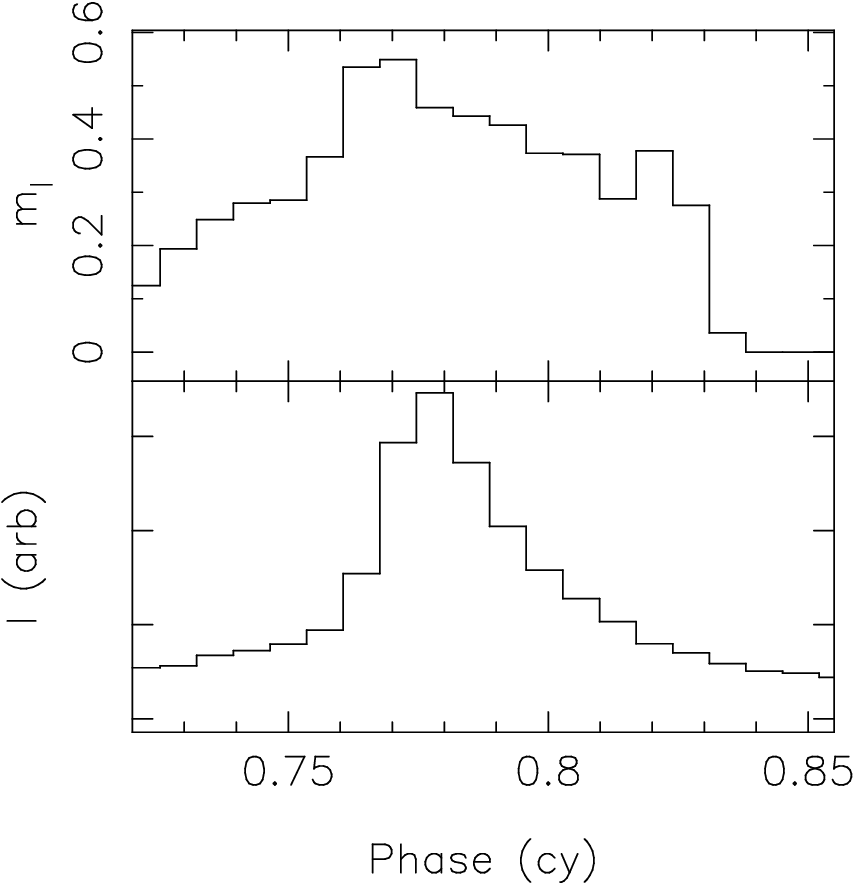}
\caption{ \label{fig:mindex}    Top panel: Modulation index  $m_I$ as a function of pulse phase $\phi$ calculated using equation (\ref{eqn:modulation}).       The leading edge of the pulse 
shows the largest $m_I$.  Bottom panel: Average pulse profile. } 

\end{center}
\end{figure}

The presence of bright single pulses on the leading edge of the integrated pulse can be verified through a number of other diagnostics.    
One of these is the intensity modulation index, defined as
\be
\label{eqn:modulation}
m_I(\phi) = \frac{\sqrt{\sigma_I^2(\phi) - \sigma_{I,{\rm off}}^2}}{I(\phi)},
\ee 
where  $I(\phi)$ and $\sigma_I(\phi)$ are the mean and rms intensity at phase $\phi$ and $\sigma _{I,{\rm off}}$ is the rms off pulse intensity. 
The modulation index  measures the 
normalized rms intensity across pulse phase $\phi$ and is sensitive 
to the frequency of occurrence of pulses at each pulse phase. 
A comparison of $m_I(\phi)$ with the integrated profile 
(Figure \ref{fig:mindex}) indicates that
the modulation index is 
largest at the leading edge of the pulse.
This suggests the existence of a tail of  
bright pulses on the leading edge that has large enough
amplitudes to dominate the modulation index.   

\begin{figure}[!ht]
\begin{center}
\includegraphics[scale=0.5]{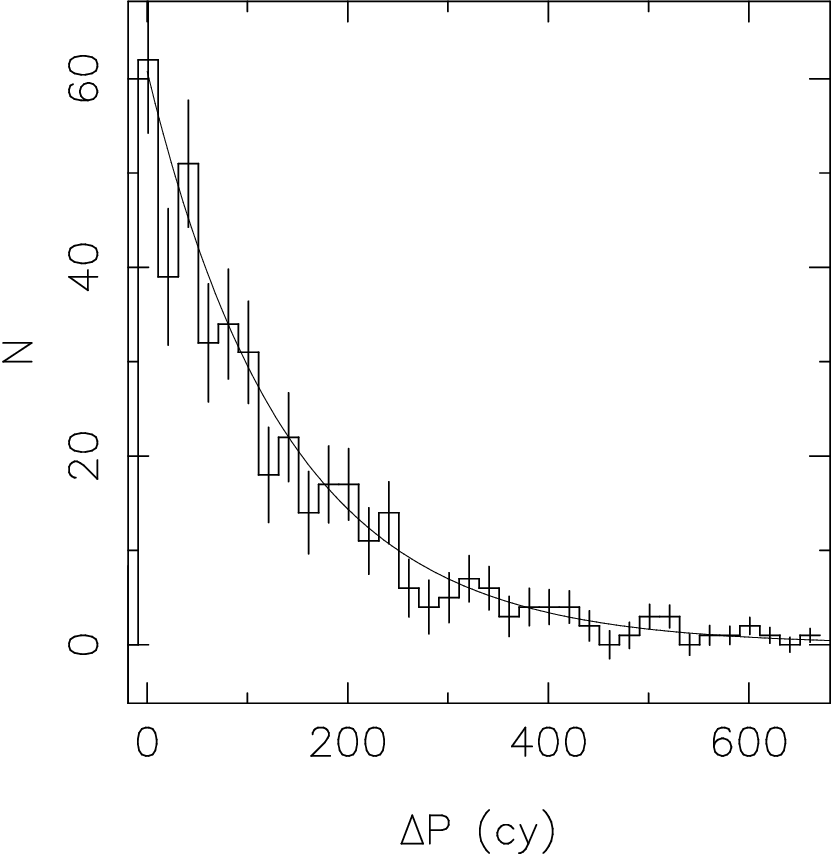} 
\caption{ \label{fig:dthist}  Histogram of the spacing  $\Delta P$ of bright pulses. The thin line shows the expected exponential distribution for 
$\Delta P$  if 
the number of bright pulses per unit time is 
  Poisson distributed.
}
\end{center}
\end{figure}

The separations between strong pulses in units of the pulse period appear
to be random. 
In Figure \ref{fig:dthist}, we show the histogram for the spacing in arrival time $\Delta P$ for bright pulses  with $S/N>14$.   The histogram of $\Delta P$ is fit well by an exponential distribution and is therefore consistent with a Poisson process.
The mean spacing is $\sim 150$~pulses.
The exponential distribution is the same as found for giant pulses from
the Crab pulsar \cite[][]{1995ApJ...453..433L}.

\subsection{Single Pulse Timing}

\begin{figure}[!ht]
\begin{center}
\includegraphics[scale=0.5]{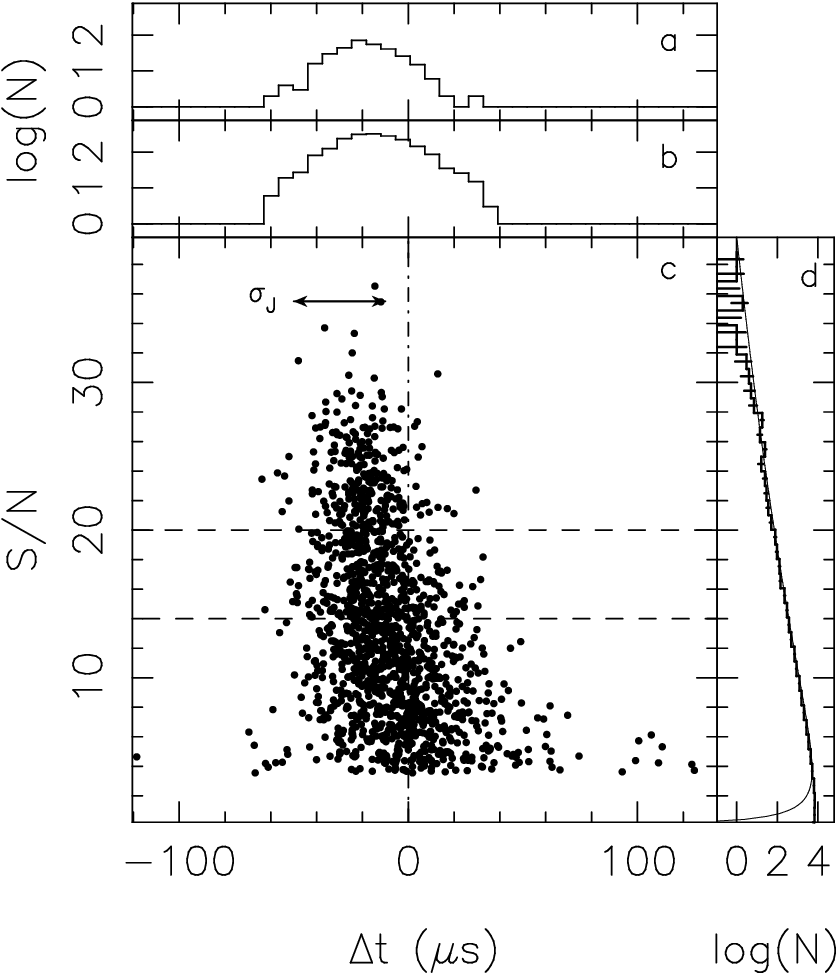}
\caption{ \label{fig:sn_vs_phase} Single pulse amplitude and  arrival time statistics for PSR J1713$+$0747.    Panel a:  Histogram of the arrival times for very bright ($S/N>20$) single pulses.  Panel b: Histogram for weaker single pulses ($14 < S/N < 20$).
  Panel c:  Scatter plot of residual arrival time $\Delta t$ versus $S/N$ for single pulses.  The horizontal dashed lines indicate the
S/N ranges used  for panels a and b. The vertical dashed-dot line shows the mean pulse arrival time.  For clarity, only a random selection of the single pulses have been plotted. To compensate for fewer bright single pulses, 
the probability of a pulse being selected was scaled $\propto 1 /\rho(S/N_i)$, where $\rho(S/N)$ is the probability density function for S/N derived empirically from panel {\em d}.    
 Panel d: Histogram of the S/N of single pulses.  On the histogram, we also plot the $1\sigma$ counting errors.  The best-fit curve to a log-normal distribution is displayed as a thin line.  
}

\end{center}
\end{figure}

There is a modest correlation between single pulse 
$S/N$ and pulse arrival time. 
In Figure \ref{fig:sn_vs_phase}, panel c, we show a scatter plot of residual arrival times $\Delta t$ versus $S/N$ for all pulses with $S/N > 3$.  
Brighter pulses tend to have earlier arrival times. 
In panels a and b, we display histograms of $\Delta t$ for bright ($S/N > 20$) 
and weak ($14 < S/N <20$) pulses, respectively.  
The weaker pulses arrive in a wide range of $\Delta t$, whereas 
the brightest pulses arrive in 
a narrower range of arrival time. 
While there is a clear correlation between arrival time and S/N, the correlation is not strong and the scatter in arrival times (labeled $\sigma_J$ on the plot) is comparable in 
magnitude of the correlation. 
We note that a correlation between $S/N$ and arrival time had previously been observed in the bright canonical 
pulsar PSR~B0329$+$54 \cite[]{1993A&A...269..325M}.  This correlation was identified with mode changing and has a different character than the fast jitter effect observed in PSR~J1713$+$0747. 

In Figure  \ref{fig:sn_vs_phase}, panel d,  we show the histogram for S/N, which is a proxy for pulse intensity.  For S/N $>$ 3.5, the distribution is  well described by a  log-normal distribution.  We fit a log-normal distribution to the histogram for S/N $>$3.5 and found a reduced $\chi^2$ of $1.7$.   This distribution is consistent with what is observed in canonical pulsars \cite[][]{2004MNRAS.353..270C}.
   The distribution diverges from a log-normal distribution when $S/N \lesssim 3$ because radiometer noise (and not intrinsic brightness) dominate the estimate of pulse intensity. 
  The distribution is highly inconsistent with a power-law distribution because the observed distribution does not show a long tail of pulses at high S/N.  The best-fit power law produced poorly matched the data and had a reduced $\chi^2$ of $150$.      

\subsection{Timing Fluctuations for Integrated Profiles}

Analysis of integrated pulse profiles produced by the ASP instrument 
confirms that intrinsic pulse shape variations are manifested as 
TOA variations in the integrated profiles.

\begin{figure}[!ht]
\begin{center}
\includegraphics[scale=0.5]{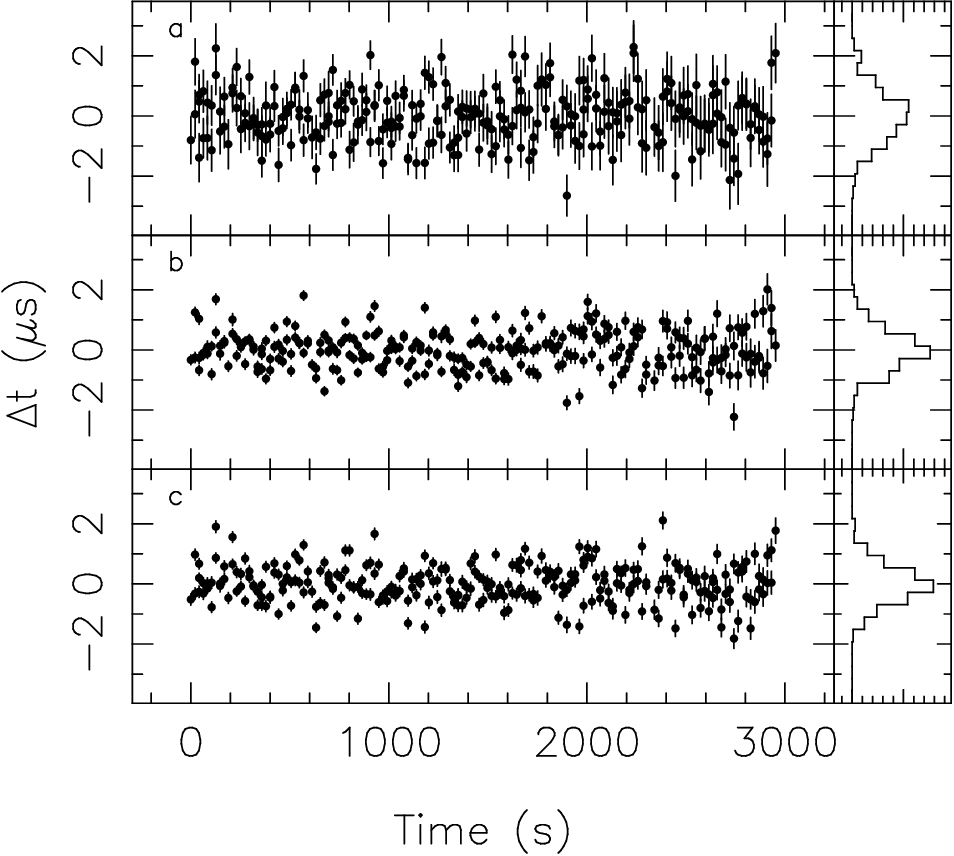}
\caption{ \label{fig:TOAs_L_S}  Left panels:  Residual TOAs for observations simultaneously obtained in three sub-bands of 4~MHz bandwidth.  The central frequencies of the bands are $1382$~MHz for panel {\em a}, $1410$~MHz for panel {\em b}, and $1430$~MHz for panel {\em c}.   
Right panel: Histograms of residual TOAs.   
} 
\end{center}
\end{figure}

In Figure \ref{fig:TOAs_L_S}, the residual TOAs for ASP observations 
are displayed for observations from MJD~$54978$, a day in 
which diffractive and refractive scintillation enhanced the 
flux in the $64$~MHz ASP band by a factor of $\approx 4$.   
Within  the figure, we show residual arrival times for 
simultaneous observations within three  sub-bands of 
$4$~MHz bandwidth, widely spaced across the band. 
The error bars represent the white noise uncertainty in  the 
TOA estimates and vary in size between the sub-bands and 
in time because the flux (though relatively high in all bands) 
is modulated by diffractive scintillation. 

Figure \ref{fig:TOAs_L_S} also shows  histograms of the residual TOAs for
the three sub-bands. 
The values for skewness and kurtosis that we estimate indicate that the
histograms  are consistent with Gaussian distributions.  

\begin{figure}[!ht]
\begin{center}
 \includegraphics[scale=0.5]{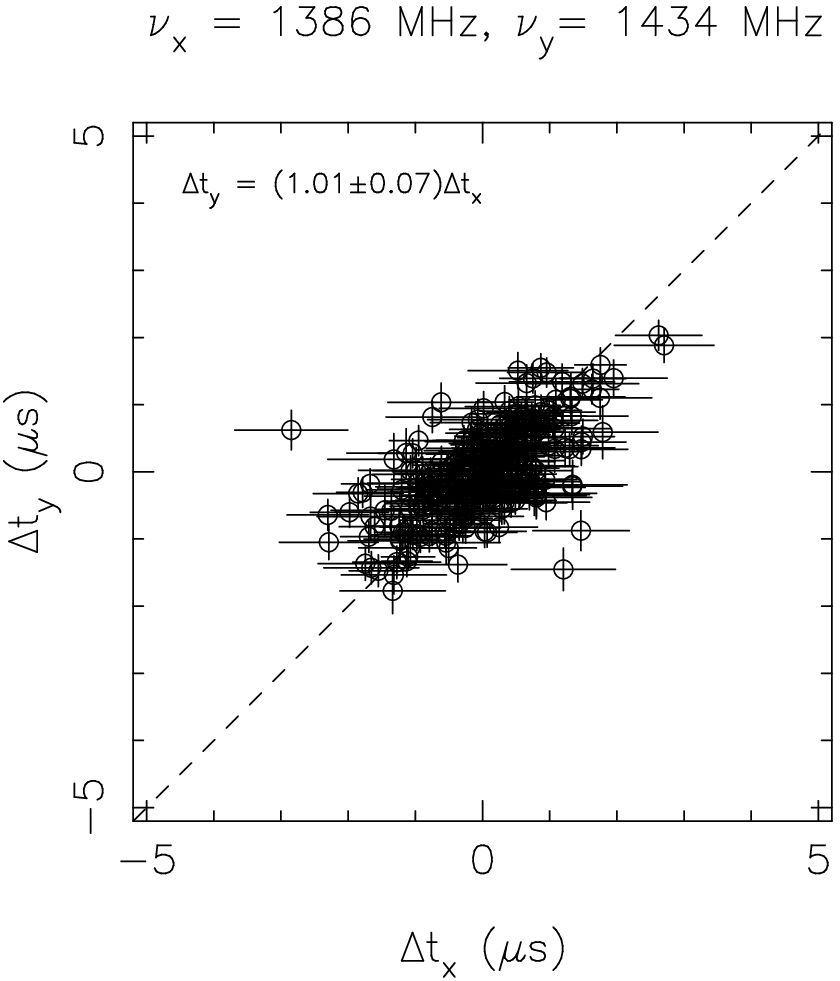}
\caption[Correlation Between TOAs with in observations of J1713$+$0747.]{ \label{fig:corr_L}  Correlation of TOAs between  two frequency channels.  The error bars represent the $1-\sigma$ template fitting errors.  The dashed angled line indicates $\Delta t_y = \Delta t_x$.     } 
\end{center}
\end{figure}

The arrival times are strongly correlated between the sub-bands.  In  Figure \ref{fig:corr_L}, we show a scatter plot of the residual arrival times in two frequency bands.   A best fit line to the data shows a slope consistent with unity, implying that the 
white-noise component of the TOA variations not caused
by radiometer noise is essentially independent of frequency. 
This analysis was conducted for all pairs of frequency channels, and in all cases a slope consistent with unity was found.
We note that a similar correlation between sub-banded TOAs was found in Parkes observations of PSR~J0437$-$4715 by \cite{2011arXiv1108.0812O}, suggesting that the effect is not observatory dependent.   

\begin{figure}[!ht]
\begin{center}
\includegraphics[scale=0.5]{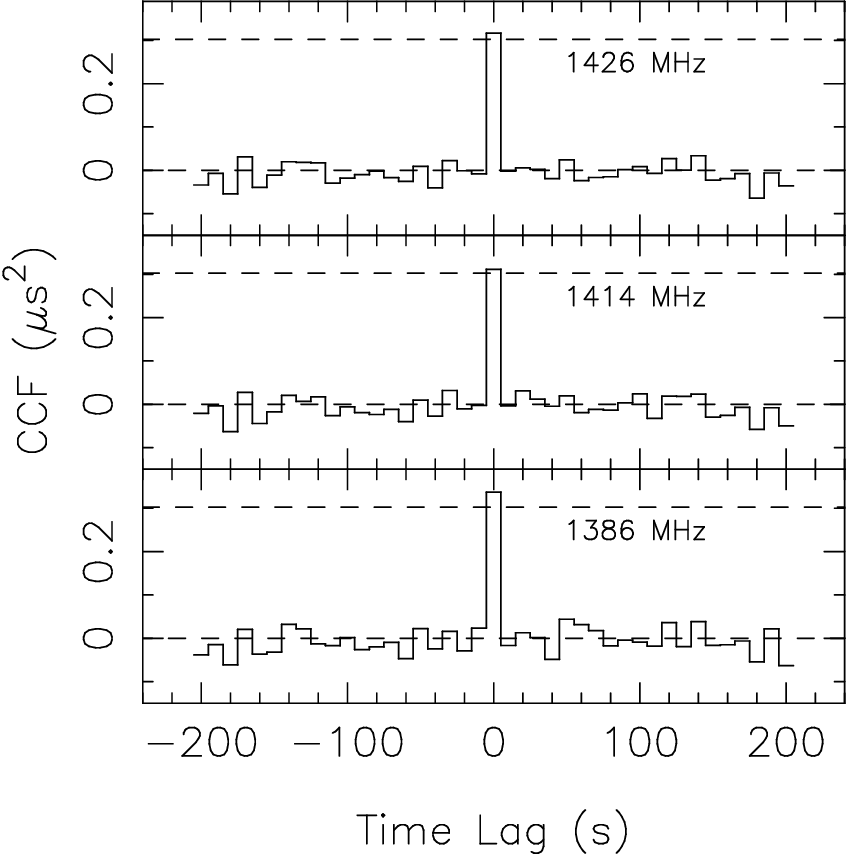} 
\caption[Cross Correlation Function of Time Series]{ \label{fig:CCF_highsnr}  Cross correlation function between time series for the $1430$~MHz channel 
and three other channels.        The lower dashed lines show the
zero correlation level.  The CCF shows a spike at zero lag that decays
to zero in one sample (10~s).  The upper dashed lines shows the level of jitter 
inferred from Figure \ref{fig:jitter_analysis}.
}
\end{center}
\end{figure}

Another demonstration of the large correlation of the TOA variations 
across frequency is given in 
 Figure \ref{fig:CCF_highsnr}, which shows the cross correlation functions 
(CCFs) for arrival time variations  
between the three pairs of  sub-bands. 
 The CCFs show  spikes at zero lag that drop to zero correlation after one
sample (10~s).   
  The zero-lag spike confirms that the residual TOAs are correlated between different frequency channels, and that the structure is unresolved in time at the $10$~s level.  
  This is consistent with what would be expected if the single pulse modulation affecting the WAPP observations was causing the timing error observed in the ASP observations, because the variations appear to be uncorrelated with time.

If we define the rms pulse-phase jitter $\sigma_J$ as the excess white-noise
residual over that expected from radiometer noise, we can estimate it
from the CCF because radiometer noise will not correlate between different
frequency bands while the jitter will.   The zero-lag value is
${\rm CCF(0) = \sigma_J^2 \approx 0.30~\mu s^2}$, so ${\rm \sigma_J \approx 0.55~\mu}$s
for $T=10$~s (2188 pulses). This corresponds to jitter in single-pulse arrival
times of $\sqrt{2188}\times 0.55~\mu s \approx 26~\mu s$,  compared to
the pulse width (FWHM) of 110$~\mu s$.
 Our observation of shape variations are consistent with a model presented in \cite{1985ApJS...59..343C}, in which  single pulses are modeled to have narrower profiles than the mean profile,  but large variations in centroid.
Shape variations  can then be characterized by a dimensionless 
jitter parameter $f_J = 1-(w_i/w_a)^2$, where $w_i$ is the width of a single pulse and $w_a$ is the width of an average pulse.   
For PSR J1713$+$0747, we find a jitter parameter $f_J \approx 0.24$, which is consistent with values found for other pulsars \cite[][]{1985ApJS...59..343C}.


 \subsection{Connecting Single Pulse Timing Variations to Standard 
Precision Timing Observations}\label{sec:connect_asp_wapp}

Phenomenologically, it has been known since the early days of pulsar
timing that despite the phase jitter of individual pulses,
 integrated profiles converge toward an  average shape that is stable
on time scales of decades.  
This convergence corresponds to timing precision that improves
as $1/\sqrt{N_p}$ when $N_p$ pulses are summed 
\cite[][]{1975ApJ...198..661H}.  
For some long-period pulsars, the effective number of independent pulses 
is less than $N_p$ because there are short term correlations and the results for J1713+0747 are consistent with a lack of correlations.

\begin{figure}[!ht]
\begin{center}
\includegraphics[scale=0.6]{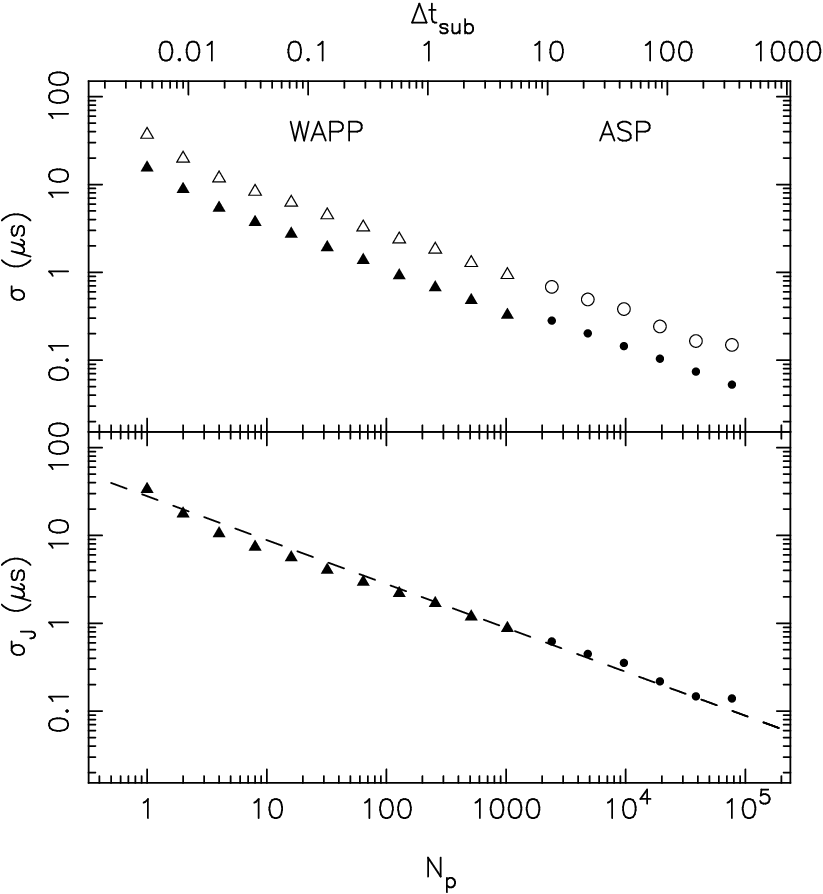} 
\caption[Residual Error Caused by Jitter in PSR~J1713$+$0747.]{ 
\label{fig:jitter_analysis}  Top panel: Measured rms (unfilled symbols) and 
expected rms (filled symbols) timing residuals  versus the number 
of pulses averaged 
$N_p$, for observations using the WAPP (triangles) and ASP backends (circles).  
Bottom panel:  Quadrature difference between observed and expected 
rms timing errors. The dashed line is a fit to the data, 
showing the expected scaling 
$\sigma_J(N_p) \propto N_{p}^{-1/2}$.  
Despite having different levels 
of radiometer noise, the level of jitter extrapolated from 
the WAPP observations is consistent with the ASP observations.  
The horizontal scale at the top of the plot 
gives the  integration time in seconds of sub-integrations  
$\Delta t_{\rm sub}$ corresponding to $N_p$ in the bottom scale.} 
\end{center}
\end{figure}

We demonstrate that the 
arrival-time errors from standard template fitting to average
profiles obtained by averaging $N_p\gg 1$ pulses are completely
consistent with the directly measured pulse-to-pulse variations. We do
so by comparing the actual timing errors from ASP data with
the errors expected under idealized conditions where an average profile
is the sum of the template and additive white noise.  
%
 In the top panel of Figure \ref{fig:jitter_analysis}, 
the rms residual $\sigma(N_p)$ is plotted against 
the number of pulses averaged, $N_p$, for both ASP and WAPP observations 
(open symbols).
We also show rms timing residuals $\sigma_r(N_p)$ 
expected if there were only additive radiometer noise and no phase
jitter (filled symbols). 
Values for $\sigma_r$  were estimated from simulated pulse profiles 
calculated 
  by adding white noise to the template shape using the S/N
appropriate for the data. 
The simulated profiles were
analyzed identically to the observations to produce arrival time 
estimates.  
These simulated profiles also satisfy the conditions for matched
filtering and therefore yield the smallest possible timing error under the assumption that there is minimally stationary (white) noise present in the timing observations \cite[][]{1057571}.
The predicted rms error for additive noise alone
is much less than $\sigma$.  
Both $\sigma$  and $\sigma_r$ decrease proportionally to $1/\sqrt{N_p}$,  
as expected for timing errors that are uncorrelated between pulses.

In the bottom panel of Figure  \ref{fig:jitter_analysis}, we show the quadrature difference between the actual and predicted timing errors
\be
\sigma_J(N_p)  = \sqrt{\sigma^2(N_p) - \sigma_r^2(N_p)}.
\ee
After accounting for the different numbers of pulses averaged, the 
values of $\sigma_J$ obtained from the WAPP and ASP observations 
consistent with each other.   
The single-pulse value $\sigma_J(1) \approx 27\pm 1~\mu$s
is consistent with the value derived from the cross-correlation analysis in the previous section.  
The total timing error is a combination of both phase jitter and
radiometer noise.  



\section{Confirming the Jitter Hypothesis} \label{sec:confirm_jitter}
  
 We have shown that the excess timing error 
is broadband and uncorrelated between
pulse profiles calculated from disjoint sets of pulses. 
   In addition to jitter, there are two other plausible forms of timing error that can contribute errors of this nature:    diffractive 
interstellar scintillation and observation calibration.   
Here we rule out both of these alternative explanations. 

Diffractive interstellar scintillation (DISS) will generally
modulate the pulse shape on minutes to hour time scales, depending
on observing frequency, dispersion measure, and direction. 
The shape perturbation is  correlated over a time equal
to the diffractive time scale $t_d$ and over a frequency range equal
to the scintillation bandwidth,  
$\nu_d$ \cite[][]{1990ApJ...349..245C,2008ApJ...674L..37H,cs2010}. 
The DISS time scale and bandwidth vary 
strongly with observing frequency, approximately as
 $\nu^{-1}$ and $\nu^{-4}$, respectively. 
The DISS timing perturbation
has been observed in another  (but much more strongly scattered) MSP, PSR~B1937$+$21 \cite[][]{1990ApJ...349..245C,2004ApJ...602L..89J}. 

DISS is ruled out as the source of the excess TOA variation
 observed for MSP J1713+0747.
First, diffractive TOA variations would be correlated over a time
comparable to the $1$~hr diffractive time scale whereas the TOA variations appear to vary on a time scale shorter than the rotation period of the pulsar ($4.6$~ms).
Also, the timing perturbation expected from DISS is much 
smaller than the observed timing error.  Diffractive scintillation
is expected to contribute an rms error of \cite[][]{1990ApJ...349..245C,2010ApJ...717.1206C,cs2010}
\be
\label{eqn:diss_rms}
\Delta t_{\rm DISS} \approx \tau/\sqrt{N_s}.
\ee  
where $\tau$ is the pulse broadening time and $N_s$ is the number of ``scintles,''
which are the number of bright patches of constructive interference contained in 
the frequency-time plane of an observation. 
The number of scintles is $N_s \approx (1+0.3 T/t_d) (1 +0.3 \Delta \nu /\Delta \nu_d)$ \cite[][]{1990ApJ...349..245C}, where $T$ is the observing time for each sub-integration (which for our observations vary in duration between the rotation period of a pulsar $4.6$~ms and $\approx 10^2$~s), 
 $t_d$ is the diffractive time scale ($> 1$~hr), $\Delta \nu$ is the observing bandwidth $100$~MHz and $\Delta \nu_d$ is the diffractive scintillation bandwidth ($\approx 50$~MHz).  
 Thus for the observations reported here, $N_s \approx 1$. 
The pulse broadening time can be inferred from the scintillation bandwidth $\tau = C_1 / 2 \pi \Delta \nu_d$, where $\Delta \nu_d$ is the scintillation bandwidth and $C_1$ is a constant of order unity \cite[][]{1998ApJ...507..846C}. 
Evaluating Equation (\ref{eqn:diss_rms}), we find
that $\Delta t_{\rm DISS} \approx 3$~ns, which is much less than the rms scatter of 
26~$\mu s$. 


It is also possible that instrumental effects
(including both calibration errors and radio frequency interference) can cause 
pulse shape changes. 
However we disfavor this origin. 
We convincingly connect the levels of jitter observed in ASP observations with that observed in WAPP observations.
These backends operate with different architectures.
They also have different sampling levels, suggesting that artifacts of digitization, such as scattered power \cite[][]{2001A&A...378..700K},  are not grossly affecting the 3-level sampled WAPP observations.
  The jitter is observed at modestly different frequencies 
(ruling out narrow-band RFI).
In a previous observing campaign in which PSR~J1713$+$0747 was observed simultaneously with the Arecibo and Green Bank telescopes, residual TOAs were found to be correlated on similarly short time scales at the same observing frequency \cite[][]{2010AAS...21545324L}.   Calibration errors associated with non-orthogonality of the feeds should be uncorrelated between the different telescope feeds.  Additionally, timing errors induced will change slowly as the orientation of the feed with respect to the pulsar changes.   
Radio frequency interference  (RFI) could also plausibly cause timing error.   Interference is local to each telescope and will be uncorrelated 
between telescopes separated by thousands of kilometers.
  
The correlation between larger S/N and earlier arrival times is seen only at the single-pulse
level and is therefore only observed in the WAPP data sets. Our interpretation is that  the effect is evidently diminished in averages of 10~s or longer.  
There is no evidence that the correlation is caused by instrumental effects.
Firstly, the effects of digitization are small because the signal to noise ratio of the detected, non-dispersed signal  is low.
  The dispersion smearing across the band is approximately $3$~ms (much greater than the pulse width) so the non-dispersed signal has an S/N of  approximately $0.7$.  
Distortions in pulse shapes are strongest when the dispersive delay across the band is comparable to the pulse width and are suppressed when the delay is much larger than the pulse width \cite[][]{2000MNRAS.314..459S}.
The dominant effect of digitization is the appearance of negative detected power (dips) at the edges of the pulse profile \cite[][]{1998PASP...110..1467J}.   We see no evidence for this type of distortion in our data.   
For example in Figure \ref{fig:sn_budget}, the baseline is flat at the edge of even the brightest pulses ($S/N > 10$) that would be the most affected by distortions.  
The most convincing demonstration of an intrinsic origin for the correlation between S/N and arrival time would be to observe the same effect with different instrumentation that provides fast-sampled data having more bits
of representation (e.g., $\geq 8$ bits) and preferably different architecture (e.g., a baseband recorder or a polyphase filterbank).

\section{Correcting TOAs for Pulse Shape Changes} \label{sec:correction}

We investigated two distinct methods for correcting arrival times due to short term variations: multi-component template fitting and principal component analysis (PCA).  These techniques were tested on both ASP and WAPP data, enabling us to test the correction algorithms on data over a wide range of pulse averaging.  
We were unable to develop an algorithm to improve the precision of 
arrival times using pulse shape information.  We attribute the 
lack of success to the absence of a sufficiently  
strong correlation between pulse shape changes and arrival times presented 
in the previous sections.      
Here we briefly summarize  two attempts at correction and defer further discussion to a future paper. 

We implemented  a multi-component TOA fitting algorithm similar  to one  that was used to improve the timing precision of PSR~J1022$+$1001 \cite[][]{1999ApJ...520..324K}.  For PSR~J1713$+$0747, we constructed an analytic template containing $5$ Gaussian components, consistent with previous models of the pulsar \cite[][]{1998ApJ...501..270K}.  We then made a  non-linear fit
for the shape parameters, i.e., the amplitude, widths, and relative positions of the different components.  The TOAs generated with this technique were then compared to TOAs inferred using the standard method.  We found no improvement in the residual arrival times from TOAs derived from the non-linear technique.

We also conducted principal component analysis (PCA) on our 
data using a method comparable to one presented in  
\cite{2011arXiv1108.0812O}, though implemented using a different algorithm.  


To briefly summarize our algorithm, we calculated data vectors by subtracting the average
pulse profile from individual profiles.   The covariance matrix of the data vectors was then
diagonalized to obtain the eigenvalues and eigenvectors.

   A scalar quantity, 
formed by projecting the observed pulse profiles onto the most  significant 
eigenvectors, is then calculated and  correlated with 
the residual TOAs.  Any significant correlation between the 
scalar and residual TOAs indicates that a particular 
eigenvector contains aspects of profile variation that 
is influencing arrival times. \cite{2011arXiv1108.0812O} 
found such an eigenvector for PSR~J0437$-$4715 and were able 
to use the correlation to improve the timing precision of the pulsar.
For PSR J1713$+$0747, we found no significant correlation 
between any eigenvector and residual arrival time and 
were therefore unable to implement any correction technique using TOAs.  

For both PSR~J1022$+$1001 and PSR~J0437$-$4715, correlated pulse shape changes are highly significant in integrated pulse profiles, with sub-integrations separated by minutes to days showing differences in  pulse shape visible by eye.  For PSR~J1713$+$0747, the pulse shape changes appear to decorrelate on pulse to pulse time scales on the millisecond-hour baselines probed by the observations presented here.

\section{Discussion and Conclusions}\label{sec:discuss}

We have demonstrated that pulse jitter is the dominant source of timing error in the observations of PSR~J1713$+$0747 presented here.
  This pulsar is currently being monitored by the three major pulsar timing array collaborations:  the European Pulsar Timing Array \cite[EPTA, ][]{2010CQGra..27h4014F}, the North American Nanohertz Observatory for Gravitational Waves \cite[NANOGrav, ][]{2009arXiv0909.1058J},  and the Parkes Pulsar Timing Array \cite[PPTA, ][]{2010CQGra..27h4015V}.
Based on the observations presented here, the timing error induced by shape variations can be extrapolated to the longer observing spans typically used in pulsar timing array observations.      
For a $30$~min observation with $N_p \approx 10^{5.6}$, 
we predict $\sigma_J \approx 40$~ns.    This is consistent with the rms error observed in the long term timing precision of this object \cite[][]{2012arXiv1201.6641D}, suggesting that a significant component of the timing error is associated with jitter.   
 
Pulse shape changes, if uncorrelated on time scales longer than typical observing cadences and uncorrectable using information about pulse shape changes, contribute additional white noise to TOAs.    
In general there are three avenues for reducing white noise in timing observations:  observing with a more sensitive telescopes, observing with a wider observing bandwidth, or averaging over a larger number of pulses.   
Of these three options,  only the last can be used to 
reduce timing error associated with jitter because the TOA perturbations are broadband and independent of antenna gain. 
 This implies that many pulsars may need to be observed with higher throughput (i.e, with longer observations at each epoch, or observed with shorter cadence) to make a convincing detection of the gravitational wave background \cite[][]{2011arXiv1106.4047C}. 

The timing precision of PSR~J1713$+$0747 and probably other {\em bright} MSPs is likely limited by jitter if observed with telescopes where the single pulse $S/N \gtrsim 1$. 
If we assume that  the pulse is Gaussian shaped with width $w$ the rms associated with jitter is \cite[][]{1985ApJS...59..343C,cs2010}
\be
\label{eqn:jitter_rms}
\Delta t_J \approx w_i f_J (1+ m_I^2) N^{-1/2},
\ee
where $w_i$ is the intrinsic pulse width, $f_J$ is the jitter parameter, $m_I$ is the modulation index, and $N$ is the  number of pulses used to form each subintegration.
In contrast,  the rms expected from radiometer noise is 
\be
\label{eqn:radio_rms}
\Delta t_R  \approx  \frac{w_{\rm a}}{S_0\sqrt{N}}
\ee
where $w_a$ is the width of the integrated pulse and  $S_0$ is the single pulse S/N.  
We can solve for the critical $S_0$ by equating equations (\ref{eqn:jitter_rms}) and (\ref{eqn:radio_rms}).
We find that the threshold single pulse signal to noise ratio is
\be
S_c \approx && \frac{w_a}{w_i f_J (1+m_I^2)} \\ \nonumber
 &&\approx \frac{1}{f_J(1-f_J)^{1/2}(1+m_I^2)  } \\ \nonumber 
&& \approx 1.4 \left( \frac{1-f_J}{0.5} \right)^{-1/2} \left(\frac{f_J}{0.5}\right)^{-1} \left(\frac{1+m_I^2}{2} \right)^{-1}.
\ee
The exact value of the transition depends on the magnitude of phase jitter, the modulation index, and the shape of the pulse.  However we expect that for most pulsars the critical single pulse S/N will be close to unity. 

Large telescopes  provide other avenues for improving TOA precision. 
First, higher S/N observations will be able to better diagnose the pulse shape changes in a larger number of MSPs.    Additionally, larger interferometric telescopes such as the SKA can provide a higher timing throughput if they possess the ability to form multiple sub-arrays.  In this case, multiple pulsars can be observed simultaneously, with the gain of each sub-array tailored to individual pulsars. 

Timing errors associated with shape changes are correctable only if they are accompanied by  gross changes in the shape of the pulse.
If not, correction is likely minimal because other astrophysical effects will cause similar displacement of the pulses (albeit on longer time scales).  
Though unsuccessful in the analysis of J1713$+$0747, multi-component template fitting and  PCA are sensitive to subtle changes in the shape of pulses.   

It is imperative to assess profile stability  in all current and 
candidate MSPs for pulsar timing arrays.  
As we have demonstrated, 
single pulse measurements are  not required to determine 
the amount of pulse phase jitter.  It can readily be inferred from 
integrated profiles.   

\acknowledgements

We thank the referee for suggestions that improved the quality of this paper. 
We thank G.~Hobbs, S.~Oslowski, and W.~van~Straten for useful discussions; 
and P.~Freire, A.~Venkataraman, and in particular T.~Ghosh for assistance 
with the observations.  RMS gratefully acknowledges NAIC for travel support 
for conducting the observations used in this analysis.  
This work was supported at Cornell University by the National Science Foundation through grant AST~0807151 and as a subaward from West
Virginia University from NSF/PIRE Grant 0968296.


\begin{thebibliography}{49}
\expandafter\ifx\csname natexlab\endcsname\relax\def\natexlab#1{#1}\fi

\bibitem[{{Armstrong}(1984)}]{1984Natur.307..527A}
{Armstrong}, J.~W. 1984, \nat, 307, 527

\bibitem[{{Cairns} {et~al.}(2004){Cairns}, {Johnston}, \&
  {Das}}]{2004MNRAS.353..270C}
{Cairns}, I.~H., {Johnston}, S., \& {Das}, P. 2004, \mnras, 353, 270

\bibitem[{{Coles} {et~al.}(2010){Coles}, {Rickett}, {Gao}, {Hobbs}, \&
  {Verbiest}}]{2010ApJ...717.1206C}
{Coles}, W.~A., {Rickett}, B.~J., {Gao}, J.~J., {Hobbs}, G., \& {Verbiest},
  J.~P.~W. 2010, \apj, 717, 1206

\bibitem[{{Cordes} \& {Downs}(1985)}]{1985ApJS...59..343C}
{Cordes}, J.~M., \& {Downs}, G.~S. 1985, \apjs, 59, 343

\bibitem[{{Cordes} \& {Rickett}(1998)}]{1998ApJ...507..846C}
{Cordes}, J.~M., \& {Rickett}, B.~J. 1998, \apj, 507, 846

\bibitem[{{Cordes} \& {Shannon}(2010)}]{cs2010}
{Cordes}, J.~M., \& {Shannon}, R.~M. 2010, arXiv:1010.3785

\bibitem[{{Cordes} \& {Shannon}(2012)}]{2011arXiv1106.4047C}
---. 2012, \apj, 750, 89

\bibitem[{{Cordes} {et~al.}(1990){Cordes}, {Wolszczan}, {Dewey}, {Blaskiewicz},
  \& {Stinebring}}]{1990ApJ...349..245C}
{Cordes}, J.~M., {Wolszczan}, A., {Dewey}, R.~J., {Blaskiewicz}, M., \&
  {Stinebring}, D.~R. 1990, \apj, 349, 245

\bibitem[{{D'Alessandro} {et~al.}(1995){D'Alessandro}, {McCulloch}, {Hamilton},
  \& {Deshpande}}]{1995MNRAS.277.1033D}
{D'Alessandro}, F., {McCulloch}, P.~M., {Hamilton}, P.~A., \& {Deshpande},
  A.~A. 1995, \mnras, 277, 1033

\bibitem[{{Demorest}(2007)}]{2007PhDT........14D}
{Demorest}, P.~B. 2007, PhD thesis, University of California, Berkeley

\bibitem[{{Demorest} {et~al.}(2010){Demorest}, {Pennucci}, {Ransom}, {Roberts},
  \& {Hessels}}]{2010Natur.467.1081D}
{Demorest}, P.~B., {Pennucci}, T., {Ransom}, S.~M., {Roberts}, M.~S.~E., \&
  {Hessels}, J.~W.~T. 2010, \nat, 467, 1081

\bibitem[{{Demorest} {et~al.}(2012){Demorest}, {Ferdman}, {Gonzalez}, {Nice},
  {Ransom}, {Stairs}, {Arzoumanian}, {Brazier}, {Burke-Spolaor}, {Chamberlin},
  {Cordes}, {Ellis}, {Finn}, {Freire}, {Giampanis}, {Jenet}, {Kaspi}, {Lazio},
  {Lommen}, {McLaughlin}, {Palliyaguru}, {Perrodin}, {Shannon}, {Siemens},
  {Stinebring}, {Swiggum}, \& {Zhu}}]{2012arXiv1201.6641D}
{Demorest}, P.~B., {et~al.} 2012, ArXiv e-prints

\bibitem[{{Dowd} {et~al.}(2000){Dowd}, {Sisk}, \&
  {Hagen}}]{2000ASPC..202..275D}
{Dowd}, A., {Sisk}, W., \& {Hagen}, J. 2000, in Astronomical Society of the
  Pacific Conference Series, Vol. 202, IAU Colloq. 177: Pulsar Astronomy - 2000
  and Beyond, ed. {M.~Kramer, N.~Wex, \& R.~Wielebinski}, 275

\bibitem[{{Edwards} \& {Stappers}(2003)}]{2003A&A...407..273E}
{Edwards}, R.~T., \& {Stappers}, B.~W. 2003, \aap, 407, 273

\bibitem[{{Ferdman}(2008)}]{2008PhDTFermdan}
{Ferdman}, R. 2008, PhD thesis, University of British Columbia

\bibitem[{{Ferdman} {et~al.}(2010){Ferdman}, {van Haasteren}, {Bassa},
  {Burgay}, {Cognard}, {Corongiu}, {D'Amico}, {Desvignes}, {Hessels},
  {Janssen}, {Jessner}, {Jordan}, {Karuppusamy}, {Keane}, {Kramer},
  {Lazaridis}, {Levin}, {Lyne}, {Pilia}, {Possenti}, {Purver}, {Stappers},
  {Sanidas}, {Smits}, \& {Theureau}}]{2010CQGra..27h4014F}
{Ferdman}, R.~D., {et~al.} 2010, Classical and Quantum Gravity, 27, 084014

\bibitem[{{Foster} \& {Backer}(1990)}]{1990ApJ...361..300F}
{Foster}, R.~S., \& {Backer}, D.~C. 1990, \apj, 361, 300

\bibitem[{{Helfand} {et~al.}(1975){Helfand}, {Manchester}, \&
  {Taylor}}]{1975ApJ...198..661H}
{Helfand}, D.~J., {Manchester}, R.~N., \& {Taylor}, J.~H. 1975, \apj, 198, 661

\bibitem[{{Hellings} \& {Downs}(1983)}]{1983ApJ...265L..39H}
{Hellings}, R.~W., \& {Downs}, G.~S. 1983, \apjl, 265, L39

\bibitem[{{Hemberger} \& {Stinebring}(2008)}]{2008ApJ...674L..37H}
{Hemberger}, D., \& {Stinebring}, D.~R. 2008, \apjl, 674, L37

\bibitem[{{Hotan} {et~al.}(2004){Hotan}, {van Straten}, \&
  {Manchester}}]{2004PASA...21..302H}
{Hotan}, A.~W., {van Straten}, W., \& {Manchester}, R.~N. 2004, PASA, 21, 302

\bibitem[{{Jaffe} \& {Backer}(2003)}]{2003ApJ...583..616J}
{Jaffe}, A.~H., \& {Backer}, D.~C. 2003, \apj, 583, 616

\bibitem[{{Jenet} \& {Anderson}(1998)}]{1998PASP...110..1467J}
{Jenet}, F.~A., \& {Anderson}, S.~B. 1998, PASP, 110, 1467

\bibitem[{{Jenet} {et~al.}(1998){Jenet}, {Anderson}, {Kaspi}, {Prince}, \&
  {Unwin}}]{1998ApJ...498..365J}
{Jenet}, F.~A., {Anderson}, S.~B., {Kaspi}, V.~M., {Prince}, T.~A., \& {Unwin},
  S.~C. 1998, \apj, 498, 365

\bibitem[{{Jenet} \& {Gil}(2004)}]{2004ApJ...602L..89J}
{Jenet}, F.~A., \& {Gil}, J. 2004, \apjl, 602, L89

\bibitem[{{Jenet} {et~al.}(2005){Jenet}, {Hobbs}, {Lee}, \&
  {Manchester}}]{2005ApJ...625L.123J}
{Jenet}, F.~A., {Hobbs}, G.~B., {Lee}, K.~J., \& {Manchester}, R.~N. 2005,
  \apjl, 625, L123

\bibitem[{{Jenet} {et~al.}(2009){Jenet}, {Finn}, {Lazio}, {Lommen},
  {McLaughlin}, {Stairs}, {Stinebring}, {Verbiest}, {Archibald}, {Arzoumanian},
  {Backer}, {Cordes}, {Demorest}, {Ferdman}, {Freire}, {Gonzalez}, {Kaspi},
  {Kondratiev}, {Lorimer}, {Lynch}, {Nice}, {Ransom}, {Shannon}, \&
  {Siemens}}]{2009arXiv0909.1058J}
{Jenet}, F.~A., {et~al.} 2009, astro-ph/0909.1058

\bibitem[{{Kouwenhoven} \& {Vo{\^u}te}(2001)}]{2001A&A...378..700K}
{Kouwenhoven}, M.~L.~A., \& {Vo{\^u}te}, J.~L.~L. 2001, \aap, 378, 700

\bibitem[{{Kramer} {et~al.}(1998){Kramer}, {Xilouris}, {Lorimer}, {Doroshenko},
  {Jessner}, {Wielebinski}, {Wolszczan}, \& {Camilo}}]{1998ApJ...501..270K}
{Kramer}, M., {Xilouris}, K.~M., {Lorimer}, D.~R., {Doroshenko}, O., {Jessner},
  A., {Wielebinski}, R., {Wolszczan}, A., \& {Camilo}, F. 1998, \apj, 501, 270

\bibitem[{{Kramer} {et~al.}(1999){Kramer}, {Xilouris}, {Camilo}, {Nice},
  {Backer}, {Lange}, {Lorimer}, {Doroshenko}, \&
  {Sallmen}}]{1999ApJ...520..324K}
{Kramer}, M., {et~al.} 1999, \apj, 520, 324

\bibitem[{{Kramer} {et~al.}(2006){Kramer}, {Stairs}, {Manchester},
  {McLaughlin}, {Lyne}, {Ferdman}, {Burgay}, {Lorimer}, {Possenti}, {D'Amico},
  {Sarkissian}, {Hobbs}, {Reynolds}, {Freire}, \&
  {Camilo}}]{2006Sci...314...97K}
---. 2006, Science, 314, 97

\bibitem[{{Lam} \& {Demorest}(2010)}]{2010AAS...21545324L}
{Lam}, M., \& {Demorest}, P. 2010, \baas

\bibitem[{{Lattimer} \& {Prakash}(2007)}]{2007PhR...442..109L}
{Lattimer}, J.~M., \& {Prakash}, M. 2007, \physrep, 442, 109

\bibitem[{{Liu} {et~al.}(2012){Liu}, {Keane}, {Lee}, {Kramer}, {Cordes}, \&
  {Purver}}]{2012MNRAS.420..361L}
{Liu}, K., {Keane}, E.~F., {Lee}, K.~J., {Kramer}, M., {Cordes}, J.~M., \&
  {Purver}, M.~B. 2012, \mnras, 420, 361

\bibitem[{{Lundgren} {et~al.}(1995){Lundgren}, {Cordes}, {Ulmer}, {Matz},
  {Lomatch}, {Foster}, \& {Hankins}}]{1995ApJ...453..433L}
{Lundgren}, S.~C., {Cordes}, J.~M., {Ulmer}, M., {Matz}, S.~M., {Lomatch}, S.,
  {Foster}, R.~S., \& {Hankins}, T. 1995, \apj, 453, 433

\bibitem[{{Lyne} {et~al.}(2010){Lyne}, {Hobbs}, {Kramer}, {Stairs}, \&
  {Stappers}}]{2010Sci...329..408L}
{Lyne}, A., {Hobbs}, G., {Kramer}, M., {Stairs}, I., \& {Stappers}, B. 2010,
  Science, 329, 408

\bibitem[{{Manchester} {et~al.}(2005){Manchester}, {Hobbs}, {Teoh}, \&
  {Hobbs}}]{2005AJ....129.1993M}
{Manchester}, R.~N., {Hobbs}, G.~B., {Teoh}, A., \& {Hobbs}, M. 2005, \aj, 129,
  1993

\bibitem[{{McKinnon} \& {Hankins}(1993)}]{1993A&A...269..325M}
{McKinnon}, M.~M., \& {Hankins}, T.~H. 1993, \aap, 269, 325

\bibitem[{{Oslowski} {et~al.}(2011){Oslowski}, {van Straten}, {Hobbs},
  {Bailes}, \& {Demorest}}]{2011arXiv1108.0812O}
{Oslowski}, S., {van Straten}, W., {Hobbs}, G., {Bailes}, M., \& {Demorest}, P.
  2011, arXiv:1108.0812

\bibitem[{{Sesana} {et~al.}(2008){Sesana}, {Vecchio}, \&
  {Colacino}}]{2008MNRAS.390..192S}
{Sesana}, A., {Vecchio}, A., \& {Colacino}, C.~N. 2008, \mnras, 390, 192

\bibitem[{{Shaffer} {et~al.}(1999){Shaffer}, {Kellermann}, \&
  {Cornwell}}]{1999ApJ...515..558S}
{Shaffer}, D.~B., {Kellermann}, K.~I., \& {Cornwell}, T.~J. 1999, \apj, 515,
  558

\bibitem[{{Shannon} \& {Cordes}(2010)}]{sc2010}
{Shannon}, R.~M., \& {Cordes}, J.~M. 2010, \apj, 725, 1607

\bibitem[{{Stairs} {et~al.}(2000){Stairs}, {Splaver}, {Thorsett}, {Nice}, \&
  {Taylor}}]{2000MNRAS.314..459S}
{Stairs}, I.~H., {Splaver}, E.~M., {Thorsett}, S.~E., {Nice}, D.~J., \&
  {Taylor}, J.~H. 2000, \mnras, 314, 459

\bibitem[{{Taylor}(1992)}]{1992RSPTA.341..117T}
{Taylor}, J.~H. 1992, Royal Society of London Philosophical Transactions Series
  A, 341, 117

\bibitem[{{Taylor} \& {Weisberg}(1982)}]{1982ApJ...253..908T}
{Taylor}, J.~H., \& {Weisberg}, J.~M. 1982, \apj, 253, 908

\bibitem[{Turin(1960)}]{1057571}
Turin, G. 1960, Information Theory, IRE Transactions on, 6, 311

\bibitem[{{van Straten} \& {Bailes}(2011)}]{2011PASA...28....1V}
{van Straten}, W., \& {Bailes}, M. 2011, PASA, 28, 1

\bibitem[{{Verbiest} {et~al.}(2010){Verbiest}, {Bailes}, {Bhat},
  {Burke-Spolaor}, {Champion}, {Coles}, {Hobbs}, {Hotan}, {Jenet}, {Khoo},
  {Lee}, {Lommen}, {Manchester}, {Oslowski}, {Reynolds}, {Sarkissian}, {van
  Straten}, {Yardley}, \& {You}}]{2010CQGra..27h4015V}
{Verbiest}, J.~P.~W., {et~al.} 2010, Classical and Quantum Gravity, 27, 084015

\bibitem[{{Wolszczan} \& {Frail}(1992)}]{1992Natur.355..145W}
{Wolszczan}, A., \& {Frail}, D.~A. 1992, \nat, 355, 145

\end{thebibliography}
\setcounter{footnote}{0}

\end{document}